\documentstyle[aps,array]{revtex}

\newcommand{\be}{\begin{equation}}
\newcommand{\ee}{\end{equation}}

\begin{document}

\title{The Hidden Cost of Quantum Teleportation and Remote State
Preparation}
\author{R. Srikanth}
\address{Indian Institute of Astophysics, Koramangala, Bangalore- 34,
India}
\maketitle
\date{}

\begin{abstract}   	
The amount of information transferred during standard
quantum teleportation or remote state preparation is equal to the preparation 
information of the transmitted state, rather than 
the classical communication required by respective protocol. 
This is shown by noting that the information 
required to specify the operation that verifies the transmitted state
is identical to the preparation information, at the given level
of precision $m$ (in bits).
Depending on the resolution of the projective Hilbert space,
the preparation information can be made arbitrarily precise and hence 
indefinitely larger than the classical communication cost. Therefore, the 
classical communication
is insufficient to account for the transfer of preparation information,
which is then attributed to the Einstein-Podolsky-Rosen channel.
Some fundamental repercussions for relativistic quantum 
information processing are briefly discussed. 
\end{abstract}

~\\
Quantum information differs from the classical information in respect of
information storage and retrieval. Classically,
the information (in bits) extractable from a system through measurement 
is identical to that needed to prepare/specify one of its microstates. 
On the other hand, the maximum
information that a quantum system can yield is the von Neumann
information, $S(\rho ) = -$tr$(\rho\log \rho)$, which is 1 bit for a qubit, in 
the state $\rho =
\hat{I}/2$. It never exceeds the preparation information, which 
determines the precision with which the preparatory operation
must be applied to a qubit in a reference state to produce a given 
state vector. This property of quantum information is connected to the
unpredictability of measurements, indistinguishability of non-orthogonal states
 and unclonability of unknown quantum states in quantum mechanics 
(QM) \cite{cav96}.

The amount of information required to specify an arbitrary state vector is 
clearly infinite. For our discussion, we begin by considering qubit states at
a finite precision of specification.
The chosen precision $m$ (bits per amplitude) of the prepared state
depends on how much 
we want to resolve the qubit's Hilbert space. In the geometric formulation of
the Hilbert space \cite{woo,bra94}, the minimum seperation between 
two microstates (i.e, state vectors) is given by $\phi = 2^{-m/2}$,
where $\phi$ is the smallest resolvable Hilbert space angle, a measure of 
distance given by
the Fubini-Study metric, a Riemannian metric defined on projective Hilbert
space \cite{sch94}. For a qubit specified at $m = 2$ bytes precision,
state resolution is a ``sphere" of size 
$2^{-m}\pi \simeq 4.8 \times 10^{-5}$ radians.  For a uniform
ensemble of all resolvable state vectors $|j\rangle$, 
each with probability $p_j$, on average:
\be
\label{sv}
H(\tilde{p}) \equiv - \sum_j p_j \log p_j = (D - 1)m ~{\rm bits},
\end{equation}
where $D$ is the Hilbert space dimension, 
is the preparation information required to construct an arbitrary $|j\rangle$.
For the $m$ bits required to specify a qubit state vector \cite{qubyte},
$m/2$ bits specify the real, and $m/2$ bits, the imaginary
part of the amplitude. In general,  $H(\tilde{p}) \ge S(\rho )$ 
because $S(\rho)$ measures entropy with basis states as the
statistical alternatives, whereas $H(\tilde{p})$ measures entropy with all
resolvable state vectors as the alternatives. Therefore, $H(\tilde{p}) =
S(\rho )$ when the ensemble consists only of orthogonal states.

From the viewpoint of information processing, standard quantum teleportation 
(QT) \cite{ben93} and remote state preparation (RSP) \cite{ben00} are
remarkable predictions of quantum mechanics, in that they enable
the reconstruction of an arbitrary quantum state 
at the cost of only 2 bits or possibly less (in the case of RSP) of classical 
communication. 
QT has been experimentally observed \cite{telexp}.
Even though there is room for improving the reliability of existing 
experimental realizations of QT \cite{rely}, its in-principle feasibility is 
sufficient for our discussion.
For studying the information transfer in QT, state representation in terms of
density operators is not advantageous. 
It doesn't distinguish between an {\em epistemic}
(based on what is known) and {\em ontic} (based on what actually is, if
applicable) state.  If the transmitted
state is unknown both to the sender (called Alice) and 
receiver (called Bob), then-- epistemically--
Bob's state remains the same throughout the QT. Just before the start of the QT 
of an unknown (to Alice and Bob) state $|\psi\rangle$, the reduced 
density operator, tr$_A (|\Psi^-\rangle\langle\Psi^-|)$, of Bob's
entangled particle is $\hat{I}/2$, where $|\Psi^-\rangle = 
\frac{1}{\sqrt{2}}(|\uparrow\downarrow\rangle - |\downarrow\uparrow\rangle )$.
Bob's density operator retains this value just after Alice's 
Bell state measurement, but before her classical communication
(as expected on basis of the no-signaling argument \cite{shi})
and also after her classical communication.
Thus, from the epistemic viewpoint, represented by Alice and Bob, QT has not
altered the state of Bob's qubit, and no information has passed from Alice to
Bob's qubit. 
In fact, this is related to the fact that in quantum information
probabilities arising from classical ignorance get mixed up
with "useless" information arising purely from quantum indeterminacy. 

To better visualize the information budget involved in QT, we can introduce 
the ontic viewpoint through a third party, called Charlie.
Alice and Bob share an indefinitely
large store of entangled pairs in the state $|\Psi^-\rangle$. All three 
parties agree upon a pre-arranged reference state, say $|{\rm ref}\rangle$.
Charlie prepares a pure ensemble of qubits with precision $m$ bits
and gives it to Alice, who teleports it to Bob. 

From the viewpoint of quantifying information transfer, RSP is equivalent
to the ontic aspect of QT.
Both protocols have the same goal, namely to teleport a quantum state from
Alice to Bob consuming, in the simplest instance,
one ebit (one pair of maximally entangled qubits) of entanglement. 
The difference is that in RSP, Alice transmits to Bob a {\em known} (to her) 
quantum state at a classical communication cost possibly lesser than that
required for QT \cite{pati,lo}. Therefore, 
the epistemological duality in discussing quantum
teleportation, and the concommittant need for Charlie to serve as Alice's
ontic foil, is absent in RSP.
In the context of remote state transmission, RSP 
permits us to exclude the epistemologically somewhat
elusive {\em unknown} quantum state from our discussion. 
Although we use RSP for the following discussion, we wish to point out that
one may equivalently use QT, but in this case, replacing Alice by Charlie as
the person who has complete (upto given precision) classical knowledge of the 
transmitted state.

For the preparation of the pure state, Alice either unitarily evolves
$|{\rm ref}\rangle$ with a suitable designer
Hamiltonian or selectively measures a suitable observable on 
$|{\rm ref}\rangle$. It is assumed that Alice's transmission of the
prepared state to Bob is somehow of maximum fidelity.

If the teleported pure ensemble is sufficiently large, Bob can estimate
probabilities $p_j$ of eigenstates as the observed frequency $f_j$.
For large sample size $N$, the normal probability distribution for the
frequencies is given by $\sqrt{N/2\pi p_j}\exp [(-N/2)(f_j - p_j)^2/p_j]$.
He then confirms with Alice that the estimated $p_j$'s agree with 
the state Alice prepared. For high precision preparation, this can be
prohibitively costly. A better alternative is that Alice classically
communicates her $m$ bits of preparation information 
to Bob so that the latter might {\em verify} the teleportation.
We define verifiable information as the information specifying, at $m$-bit 
precision level, the measurement on a qubit that is guaranteed to produce
some predetermined outcome with a success
probability $P_s > 1 - 2^{-m}$.
This limit is obtained by considering the overlap for the
smallest resolvable state seperation in the projective Hilbert space of 
a qubit, and Taylor expanding upto second order.

For example,
suppose Bob knows Alice to have prepared the teleported state by rotation
about the $y$-axis.  Bob can check whether it is in the state
$ |\eta\rangle = \alpha |\uparrow\rangle + \beta |\downarrow\rangle , $
where $\alpha , \beta$ are real and satisfy $\alpha^2 + \beta^2 = 1$,
by measuring the operator:
\begin{equation}
\label{oper}
\hat{M}(\theta ) = \left( \begin{array}{cc}
        \cos\theta & \sin\theta \\
        \sin\theta & -\cos\theta
        \end{array}
  \right),
\end{equation}
where $\theta = 2\cos^{-1}\alpha$ and ascertaining that $\hat{M}(\theta) = +1$. 
Equivalently, Bob
rotates the teleported state through $-\theta$, and measures 
$\hat{M}(0)$, to ascertain that the outcome is +1.
(Alternatively, Bob sends his qubits to Alice, who performs 
the tests). 
For example, at two byte precision, Alice can set $\theta = 44.8881^{\circ}$
and classically communicate this value or $\alpha$ to Bob. 
To see that all the $m$ bits are significant to the teleportation, let's 
suppose the last $n$ bits in the preparation information are 
ignored during verification. The 
resulting resolution of Hilbert space is coarser by a factor
$2^{n/2}$ and the probability of verification falls to 
$P_s > 1 - 2^{-(m - n)}$. Thus, the 
number of measurements that don't verify the prepared state increases by a 
factor $2^{n}$. 

Because Bob can verify the transmission to the given accuracy, Alice
and Bob know that
RSP indeed transfers the $m$ bits of preparation information.
Since only 2 bits (in general, 2$S(\rho)$ bits) per state
are communicated classically in QT and possibly even less in RSP, 
the classical channel
is clearly insufficient to account for the transfer of the
preparation information. Hence, the Einstein-Podolsky-Rosen (EPR) channel
is invoked to account for the complete information transfer. One
might suppose that the information transmitted via the EPR channel is on
average $m \pm c$ bits per run, where $c$ is the asymptotic classical
communication cost for QT or RSP. 

In case of $m - c$, the remaining c bits 
would be furnished by the classical communication. In fact, if $m = c$ then
no information
transfer would be required via the EPR channel, which would then serve 
simply as a sort of subtle reference state provider. 
Yet, from the viewpoint of state preparation,
it would imply that ensembles composed from no more than $2^c$ distinct 
state vectors can exist, contrary to our experience.
In case of $m + c$, the 
redundant c bits would be identified by the classical communication. In fact, 
neither case is true since the classical communication is not precision
information. Alice knows that its inclusion or exclusion does not change the 
resolution of the teleported state in comparison with the prepared state.
In the QT version, Charlie
knows that the teleported state may not be an exact copy of the prepared
state but rotated through an angle known to him on basis of Alice's measurement
outcome.
The counter-rotations Bob subsequently performs using Charlie's classical
communication do not affect the precision of the prepared state. 

The classical communication serves only to reset the possibly rotationally 
scrambled
preparation information. The (possibly scrambled) preparation information 
transmitted through the EPR channel is the hidden
cost for QT or RSP, by which quantum entanglement subsidizes remote state
transmission. Bob's local cost to implement QT or RSP 
is only $c$ bits (ignoring the cost of producing the 
$|\Psi^-\rangle$ state). Alice's cost of transfering the state is also $c$ bits 
of classical communication. 
The true cost of QT-- that of transferring the preparation information-- is 
borne by quantum entanglement and hidden from Alice's and Bob's view.

The $m$ bits of quantum information are also sufficient as the nonlocal
augmentation (i.e, the
information in bits about Alice's detector's setting given to Bob before his
measurement) required to simulate {\em any}
aspect of quantum statistics via a local realistic model of QM. However, 
interestingly, it turns out that specific non-classical manifestations,
such as the cosine correlation 
appearing in Bell inequalities \cite{bell}, can be simulated using realistic
models with finite bit nonlocality, independent of the
resolution of the underlying Hilbert space \cite{brassard,steiner}.

The transferred information is exponentially larger when we consider
the QT of continuous variables \cite{lev94}. 
Let ${\cal V}_{\phi}$ be the phase space accessible to the
particle. The number $\cal N$ of phase space cells available to the particle 
is given by $\cal N$ = 
${\cal V}_{\phi}/\hbar^3$. Unlike the classical situation, the quantum
object can exist potentially in a superposition of these phase space cell
locations. The $\cal N$ cells are the basis states,
requiring $(\cal{N}$$ - 1)m$ bits of preparation information.
 
Bell inequalities imply that no local-realistic 
theory can reproduce quantum statistics. For a non-realistic
(in the EPR \cite{EPR} sense) theory like standard quantum mechanics, it
apparently does not forbid a local character. However, this would be at
variance with the observation that in QT or RSP information is transmitted 
through the EPR channel `instantaneously' as Alice performs
her joint measurement. What this means is easily seen from Charlie's viewpoint
in a QT protocol, or from Alice's in an RSP protocol. Let's consider the latter.
It is assumed that Alice and Bob share an ebit of entanglement, and that she
proposes to transmit to him a known qubit state $|\eta\rangle$ lying on the 
equator of the Bloch sphere. Bob is far away, perhaps lightyears away! From the 
standpoint of the standard formalism of QM, Alice knows that just before 
her measurement in the basis $(\eta, \eta^{\perp})$,
Bob's particle is given by the entangled state vector
$|\Psi^-\rangle$. Just after her measurement, she knows Bob's state
vector is the disentangled state $\hat{\sigma}_z|\eta\rangle$ or 
$|\eta\rangle$ \cite{ben00}. 
She concludes that the verifiable quantum preparation information 
($\geq m$ bits) required to 
disentangle Bob's qubit was transmitted in some sense instantaneously to Bob. 
QT/RSP may be said to impose a stronger  
restriction on models reproducing quantum predictions than does the violation
of Bell inequalities, because no assumption about realism is needed for them.

The four-way uncertainty in the value of the quantum information transfered
in QT, and the corresponding need for classical communication in RSP, are 
sufficient to
anchor to relativistic causality the total information flow from Alice's to
Bob's system \cite{poetic}. However, this still leaves open the question of
how to covariantly characterize in a special relativistic setting 
the ``instantaneity" of the quantum information flow in QT and
RSP \cite{srik70}.

An interesting question is whether there is an upper bound to $m$ due to
finite resolution of Hilbert space \cite{steiner}. One can
speculate the existence of a natural finite fine-graining of Hilbert space
in analogy with classical phase space. Accordingly, this sets the lower bound
to the capacity of the EPR channel. However, the
existence or non-existence of such a bound does not affect the
present discussion. 

\acknowledgements 

I am thankful to Ms Mangala Sharma for useful discussions.


\begin{thebibliography}{100}
\bibitem{poetic} With a turn of metaphor, one might say that the
fragile raft called Peaceful Coexistence \cite{shi}, tossed violently upon the 
voluminous stream of quantum information, is anchored by a relatively
thin 2-bit classical cord to the harbor of relativistic causality.
\end{thebibliography}
\end{document}